\documentclass[twocolumn,showpacs,preprintnumbers,amsmath,amssymb]{revtex4}

\usepackage{graphicx}% Include figure files
\usepackage{dcolumn}% Align table columns on decimal point
\usepackage{bm}% bold math

\begin{document}

\title{Theory of Games on Quantum Objects}
\author{Jinshan Wu{\footnote{jinshanw@physics.ubc.ca}}}
\affiliation{Department of Physics $\&$ Astronomy,\\
University of British Columbia, Vancouver, B.C. Canada, V6T 1Z1 }

\begin{abstract}
Effect of replacing the classical game object with a quantum
object is analyzed. We find this replacement requires a throughout
reformation of the framework of Game Theory. If we use density
matrix to represent strategy state of players, they are
full-structured density matrices with off-diagonal elements for
the new games, while reduced diagonal density matrix will be
enough for the traditional games on classical objects. In such
formalism, the payoff function of every player becomes Hermitian
Operator acting on the density matrix. Therefore, the new game
looks really like Quantum Mechanics while the traditional game
becomes Classical Mechanics.
\end{abstract}

\keywords{Game Theory, Hilbert Space, Probability Theory, Quantum
Game Theory}

\pacs{02.50.Le, 03.67.-a, 03.65.Yz}

\maketitle
%\newpage
%\tableofcontents
%\newpage

\section{Introduction}
The object of Game Theory is a game, a multi-player decision
making situation, usually with conflicts between players. For
example, in a Penny Flipping Game (PFG), two players play with a
coin, say initially with $head$ state. The strategies can be used
by players are $Non-flip$ and $Flip$, which, in the language of
Physics, are operators acting on the coin. The payoff is defined
such as player $1$ wins one dollar for $head$ state after both
players applied their strategies, and lose one dollar for $tail$
state. For such static strategy games, Nash Theorem has given a
closed conclusion that at least one mixture-strategy Nash
Equilibria (NE) exists for any games. Here the NE is defined that
under such state no more players will like to change its own
strategy state, and mixture strategy is defined as a probability
distribution function (PDF) over the strategy space of every
player.

Our question is how about we replace the two-side coin here with a
$\frac{1}{2}$-quantum spin? What's the effect of this on Game
Theory? It's still a game-theory question. Players can still
choose strategies to act on the spin, although they have much more
choices now. Compared with $Non-flip$ and $Flip$, in Quantum
Mechanics, any unitary $2\times2$ matrices can be used as
operators, and $\left\{I, X, Y, Z\right\}$ are the four typical
matrices of them. Now the Game Theory must answer how to define
the strategy state for this game, how to define NE, and the
existence of NE. At last, we have to ask whether such game can be
studied within the framework of Traditional Game Theory (TGT), or
should we develop a new framework but still with the same spirit
of Game Theory? In this work, we will construct a new framework,
which can be used both TGT and the game on quantum objects, named
Quantum Game Theory (QGT).

\section{Density Matrix: language we used}

Density Matrix language for Quantum Mechanics is well known. In
Schr\"{o}dinger's Picture, a state of a quantum object is
represented by a density matrix $\rho^{q}\left(t\right)$; the
evolution is described by a unitary transformation
$U\left(t\right) \triangleq U\left(0,t\right)$ as
\begin{equation}
\rho^{q}\left(t\right) =
U\left(t\right)\rho^{q}\left(0\right)U^{\dag}\left(t\right),
\label{qm}
\end{equation}
where generally $U\left(t\right)$ is determined by $H$, the
Hamiltonian of the quantum object; and any physical quantities
should be calculated by
\begin{equation}
A\triangleq\left<\hat{A}\right> = tr\left(\hat{A}\rho^{q}\right).
\label{qmquantity}
\end{equation}
Here we want to use density matrix also to describe Classical
Mechanics, which originally is described by a PDF,
$f\left(\vec{x},\vec{p}\right)$, such as in Liouville Equation.
Now we re-express it as a density matrix as
\begin{equation}
\rho^{c} =
\sum_{x\in\Omega}f\left(x\right)\left|x\left>\right<x\right|,
\label{classicalstate}
\end{equation}
where $x$ is used to represent all general configuration
variables. In fact, even for quantum objects, this kind of states
has been used by Von Neumann in his picture of quantum
measurement\cite{von} as exclusive mixture states. Its explanation
is every sample of this state gives only one realization such as
$x^{*}$ with probability $f\left(x^{*}\right)$. This is exactly
the same meaning of the PDF. $f\left(x\right)$ normalized within
$\Omega = \left\{x\right\}$, the set of its all possible states.
By assuming
\begin{equation}
\left<x\left|\right.x^{\prime}\right> =
\delta\left(x-x^{\prime}\right),
\end{equation}
the normalizing condition for both quantum and classical density
matrix can be written as
\begin{equation}
tr\left(\rho\right) = 1.
\end{equation}
It's obvious that equ(\ref{qmquantity}) is still hold for our
classical density matrix. In fact, we even can construct evolution
equation parallel to equ(\ref{qm}),
\begin{equation}
\begin{array}{ccc}
\rho^{c}\left(t\right) & = & \sum_{x}
p\left(x\right)\left|x\rightarrow x\left(t\right)\right>\left<x
\rightarrow x\left(t\right)\right|,
\\ & = & \sum_{x}
p\left(x\right)\left(T
\left|x\right>\right)\left(\left<x\right|T^{\dag}\right)
\\
& = & T\rho^{c}\left(0\right)T^{\dag}, \label{cm}
\end{array}
\end{equation}
and also easy to show $TT^{\dag} = T^{\dag}T = I$.

However, although we unified classical and quantum description by
density matrix, but those two density matrices are different. The
one for classical object is always diagonal, while the one for
quantum object has off-diagonal elements and it's diagonal only
under one special basis. This difference roots in the
non-commutative relation between quantum operators. The way to use
density matrix to describe classical objects is just like to use
complex number to reexpress expressions of real numbers. However,
when we want to unify expressions of real number and complex
number, of course, we need to work in the field of complex number.
Here, we are in the same situation: unification of description of
states both classical and quantum objects.

Not only states of classical and quantum objects, operators on
classical and quantum objects can also be described by Hilbert
space and density matrices. We call this the density matrix
formalism for Quantum Operators. For the unitary operators on a
$\frac{1}{2}$-spin, we know they are $2\times2$ matrices, and
generally can be expanded by $\left\{I,X,Y,Z\right\}$
as\cite{qubit}
\begin{equation}
\begin{array}{ccc} U & = & e^{i\alpha}\left(\cos{\frac{\gamma}{2}}\cos{\frac{\beta
+\delta}{2}}I + i\sin{\frac{\gamma}{2}}\sin{\frac{\beta
-\delta}{2}}X \right.\\ &&\left. -
i\sin{\frac{\gamma}{2}}\cos{\frac{\beta -\delta}{2}}Y -
i\cos{\frac{\gamma}{2}}\sin{\frac{\beta
+\delta}{2}}Z\right)\end{array}. \label{decompose}
\end{equation}
Now we regard this expansion as an decomposition of a vector $U$
under the basis of $\left\{I,X,Y,Z\right\}$ of a Hilbert space of
operators $\mathcal{H}^{*} \triangleq \left\{U\right\}$.
Fortunately, $\mathcal{H}^{*}$ is a Hilbert space with a natural
defined inner product. The summation and number product of vector
is naturally fulfilled by the corresponding usual operation on
matrices, the inner product is defined by
\begin{equation}
\left<A\right|\left.B\right>\triangleq\left(A,B\right) =
\frac{Tr\left(A^{\dag}B\right)}{Tr\left(I\right)}.
\label{inner}
\end{equation}
For simplicity, later on, $\mathbb{B}$ and $\mathbb{E}$ is used to
denote the basis, $\mathbb{B}\left(\mathcal{H}^{*}\right)$, and
the space expanded by a basis, $\mathcal{H}^{*} =
\mathbb{E}\left(\mathbb{B}\left(\mathcal{H}^{*}\right)\right)$
respectively.

Since $]mathcal{H}^{*}$ is also a Hilbert space, we can use
density matrix to represent its vectors, which now, physically, is
operators. Now, we have prepared everything we will need, the
density matrix of any operators, and any probability combinations
of operators, can be generally defined as
\begin{equation}
\rho^{op} =
\sum_{\mu,\nu\in\mathbb{B}\left(\mathcal{H}^{*}\right)}\rho^{op}_{\mu\nu}\left|\mu\right>\left<\nu\right|.
\label{opstate}
\end{equation}
This can be used as strategy state for games on both classical and
quantum operators.

\section{Density Matrix Formalism for TGT}
\label{sectgt}

First, we want to put the TGT into density matrix formalism, which
means to put the strategy states as density matrices, payoff
function as Hermitian operators, and their relation should obey
equ(\ref{qmquantity}). As we know the general mixture strategy
state of a player $i$ in TGT a PDF over $i$'s strategy space, so
the density matrix form is
\begin{equation}
\rho^{c,i} =
\sum_{\mu\in\mathbb{B}\left(\mathcal{H}^{*,c}\right)}\rho^{c,i}_{\mu\mu}\left|\mu\right>\left<\mu\right|,
\label{singletgtstate}
\end{equation}
and the density matrix of all players in a non-cooperative game is
\begin{equation}
\rho^{c,S} =\prod_{i}\otimes\rho^{c,i} \label{wholetgtstate}.
\end{equation}
Or if we denote $\left|\vec{\mu}\right> = \left|\mu_{1}, \dots,
\mu_{i}, \dots, \mu_{N}\right>$, then,
\begin{equation}
\rho^{c,S}
=\sum_{\vec{\mu}}\left(\prod_{i}\rho^{c,i}_{\mu_{i}\mu_{i}}\right)\left|\vec{\mu}\right>\left<\vec{\mu}\right|.
\label{tgtstate}
\end{equation}
The payoff matrix of player $i$ is defined as
\begin{equation}
H^{i}
=\sum_{\vec{\mu}}G^{i}\left(\vec{\mu}\right)\left|\vec{\mu}\right>\left<\vec{\mu}\right|,
\label{tgtpayoff}
\end{equation}
where $G^{i}$ is the traditional payoff function in TGT, which
give a real number when all the strategies of every player are
given, $G^{i}\left(\vec{\mu}\right)$. It's easy to check in this
abstract form, the payoff is given by
\begin{equation}
E^{i} =Tr\left(\rho^{c,S}H^{i}\right), \label{tgtpayoffvalue}
\end{equation}
where $Tr\left(\cdot\right)$ is the trace over strategy state
space, $\mathcal{H}^{*}$.

One important character should be noticed that both above
$\rho^{S}$ and $G^{i}$ have only diagonal terms, which is a
character of classical systems. It's easy to check that every
classical game can be re-expressed in this language of density
matrix and Hermitian Hamiltonian. The only difference between this
TGT and Classical Mechanics is that here every player has its own
Hamiltonian, while in Physics, we only have a common one for the
whole system. This reflects the conflict of interests between
players. For example, PFG, which in TGT notation is
\begin{equation}
G^{1,2} = \left[\begin{array}{ll}1,-1 & -1,1 \\ -1,1 & 1,-1
\end{array}\right],
\label{oldpfq}
\end{equation}
can be redefined as
\begin{equation}
H^{1,2} = \left[\begin{array}{llll}1,-1
& 0 & 0 & 0 \\ 0 & -1,1 & 0 & 0 \\ 0 & 0 & -1,1 & 0 \\0 & 0 & 0 &
1,-1
\end{array}\right],
\label{pfgpayoff}
\end{equation}
and
\begin{equation}
\rho^{c,S} = \rho^{c,1}\otimes\rho^{c,2} =
\left[\begin{array}{ll}\rho^{c,1}_{nn} & 0
\\0 & \rho^{c,1}_{ff}
\end{array}\right]\otimes\left[\begin{array}{ll}\rho^{c,2}_{nn} & 0
\\0 & \rho^{c,2}_{ff}
\end{array}\right].
\label{pfgstate}
\end{equation}

\section{Density Matrix Formalism for QGT}
\label{secqgt}

In order to put QGT into density matrix form, first, we need to
define the traditional payoff function
$G^{i}\left(\vec{\mu}\right)$, and then similarly define $H^{i}$
from $G^{i}$. However, the definition of $G^{i}$ is not trivial,
because there are infinite number of strategies (unitary
operators) even for the $\frac{1}{2}$-spin SFG. Fortunately, the
inherent relation between quantum operators, such as
equ(\ref{decompose}), will save us out of this mud. The idea is
choose a basis, $\mathbb{B\left(\mathcal{H}^{*}\right)}$ for
Hilbert space of operators $\mathcal{H}^{*}$, then define $G^{i}$
as a function over $\mathbb{B\left(\mathcal{H}^{*}\right)}$ first,
and by equ(\ref{decompose}), at last $G^{i}$ will be defined on
the whole $\mathcal{H}^{*}$. Before we go into the detail, there
is a mine making our life not so easy: like operators in Quantum
Mechanics, $G^{i}$ will also be matrix operator over
$\mathbb{B\left(\mathcal{H}^{*}\right)}$ with off-diagonal
elements. This means usually,
$G^{i}\left(\vec{\mu},\vec{\nu}\right)\neq 0$, while in TGT,
$G^{i}$ has only diagonal elements. Let's demonstrate it by one
example. In SFG, $\left\{I,X,Y,Z\right\}$ is used as the operator
basis, and the payoff is still defined such that player $1$ gets
payoff $p^{1} = \rho^{spin}_{\uparrow\uparrow} -
\rho^{spin}_{\downarrow\downarrow} = -p^{2}$, where the states are
measured in $z$-direction. This can be written in a matrix form
that
\begin{equation}
\begin{array}{lll}
P^{1} = \left[\begin{array}{ll}1 & 0
\\ 0 & -1\end{array}\right]=-P^{2} & \text{and} & p^{i} =
tr\left(P^{i}\rho^{spin}\right). \label{qgtscale}
\end{array}
\end{equation}
And the state of the spin changes according to the strategies of
players by
\begin{equation}
\rho^{spin}_{end} =
\left(U_{2}U_{1}\right)\rho^{spin}_{initial}\left(U_{2}U_{1}\right)^{\dag}.
\label{qgtevlotion}
\end{equation}
Now is easy to check
$G^{1}\left(\left\{I,I\right\},\left\{I,I\right\}\right) =1$ as
usual as in TGT, but we have new elements such as
$G^{1}\left(\left\{Y,Y\right\},\left\{Y,Y\right\}\right) = 1$, and
even off-diagonal elements
$G^{1}\left(\left\{X,X\right\},\left\{I,I\right\}\right) = 1$.
This shows $G^{i}\left(\vec{\mu},\vec{\nu}\right)\neq 0$ so that
later on,
\begin{equation}
H^{i} = \sum_{\vec{\mu},\vec{\nu}}
G^{i}\left(\vec{\mu},\vec{\nu}\right)\left|\vec{\mu}\right>\left<\vec{\nu}\right|
\label{qgtpayoff}
\end{equation}
will also have off-diagonal elements. Not only the payoff matrix,
but also the density matrix of strategy state has off-diagonal
elements. Considering a player chooses strategy $U =
\frac{1}{\sqrt{2}}\left(X+Y\right)$, expressed under the basis,
it's
\begin{equation}
\rho^{op} = \frac{1}{2}\left(\left|X\right>\left<X\right| +
\left|X\right>\left<Y\right| + \left|Y\right>\left<X\right| +
\left|Y\right>\left<Y\right|\right),
\end{equation}
which obviously has non-zero off-diagonal elements such as
$\frac{1}{2}\left|X\right>\left<Y\right|$. Generally, the state of
players in a non-cooperative QGT is
\begin{equation}
\rho^{q,S} = \prod_{i}\otimes\rho^{q,i},
\end{equation}
where
\begin{equation}
\rho^{q,i} =
\sum_{\mu,\nu\in\mathbb{B}\left(\mathcal{H}^{*,q}\right)}\rho^{q,i}_{\mu\nu}\left|\mu\right>\left<\nu\right|.
\end{equation}
Or put in another way,
\begin{equation}
\rho^{q,S} =
\sum_{\vec{\mu},\vec{\nu}}\left(\prod_{i}\rho^{q,i}_{\mu_{i}\nu_{i}}\right)\left|\vec{\mu}\right>\left<\vec{\nu}\right|.
\label{qgtstate}
\end{equation}
And then the payoff value is given by
\begin{equation}
E^{i} =Tr\left(\rho^{q,S}H^{i}\right). \label{qgtpayoffvalue}
\end{equation}
Compare equ(\ref{qgtstate}) and equ(\ref{qgtpayoff}) with
equ(\ref{tgtpayoff}) and equ(\ref{tgtstate}), we notice that the
existence of off-diagonal elements is the difference between QGT
and TGT.

Equ(\ref{qgtscale}), the scale matrix used to assign payoff value
to each player according to the state of object and
equ(\ref{qgtevlotion}), the evolution of state of object, can also
be generalized to games on any classical and quantum objects. This
has been done in \cite{wugame}, where it is named as
``Manipulative Definition'' of game, the payoff of player $i$ is
given by a physical process changing the state of the object and a
scale to readout the end state into payoff value,
\begin{equation}
E^{i}\left(S\right) =
tr\left(P^{i}\mathcal{L}\left(S\right)\rho^{object}_{inital}\mathcal{L}^{\dag}\left(S\right)\right),
\label{mpayoff}
\end{equation}
where $S = \left(s^{1}, s^{2}, \cdots, s^{N}\right)$ is an ordered
sequence of the strategies used by all players.

The manipulative definitions of PFG and SFG is given respectively
as followings,
\begin{equation}
\begin{array}{ll} \rho^{o}_{0} = \left|+1\right>\left<+1\right| = \left[\begin{array}{ll}1 & 0 \\ 0 &
0\end{array}\right], & S^{1} = S^{2}=\left\{I,X\right\}\\
\mathcal{L}\left(s^{1}, s^{2}\right) = s^{2}s^{1}, & P^{1} =
\left[\begin{array}{ll}1 & 0
\\ 0 & -1\end{array}\right]=-P^{2},
\end{array}
\label{pennygame}
\end{equation}
and
\begin{equation}
\begin{array}{ll} \rho^{o}_{0} = \left|+1\right>\left<+1\right| =\left[\begin{array}{ll}1 & 0 \\ 0 &
0\end{array}\right], & S^{1} = S^{2}=
\mathbb{E}\left\{I,X,Y,Z\right\}
\\ \mathcal{L}\left(s^{1}, s^{2}\right) = s^{2}s^{1}, &
P^{1} = \left[\begin{array}{ll}1 & 0
\\ 0 & -1\end{array}\right]=-P^{2}
\end{array}.
\label{spingame}
\end{equation}
From these manipulative definition, it's easy to see that the only
difference coming from the strategy space, that
$\left\{I,X\right\}$ for classical object and
$\mathbb{E}\left\{I,X,Y,Z\right\}$ for quantum object. And this
difference requires the off-diagonal terms in both density matrix
of strategy state and Hermitian payoff matrices. The explicit form
of $H^{1}$, payoff matrix of player $1$ in SFG is
\begin{equation}
\left[\begin{array}{cccccccccccccccc} \scriptstyle1&\scriptstyle
&\scriptstyle &\scriptstyle 1&\scriptstyle &\scriptstyle
1&\scriptstyle -i&\scriptstyle &\scriptstyle &\scriptstyle
i&\scriptstyle 1&\scriptstyle &\scriptstyle 1&\scriptstyle
&\scriptstyle &\scriptstyle 1
\\&\scriptstyle -1&\scriptstyle i&\scriptstyle &\scriptstyle -1&\scriptstyle &\scriptstyle &\scriptstyle -1&\scriptstyle i&\scriptstyle &\scriptstyle &\scriptstyle i&\scriptstyle &\scriptstyle 1&\scriptstyle -i&\scriptstyle
\\&\scriptstyle -i&\scriptstyle -1&\scriptstyle &\scriptstyle -i&\scriptstyle &\scriptstyle &\scriptstyle -i&\scriptstyle -1&\scriptstyle &\scriptstyle &\scriptstyle -1&\scriptstyle &\scriptstyle i&\scriptstyle 1&\scriptstyle
\\\scriptstyle1&\scriptstyle &\scriptstyle &\scriptstyle 1&\scriptstyle &\scriptstyle 1&\scriptstyle -i&\scriptstyle &\scriptstyle &\scriptstyle i&\scriptstyle 1&\scriptstyle &\scriptstyle 1&\scriptstyle &\scriptstyle &\scriptstyle 1
\\&\scriptstyle -1&\scriptstyle i&\scriptstyle &\scriptstyle -1&\scriptstyle &\scriptstyle &\scriptstyle -1&\scriptstyle i&\scriptstyle &\scriptstyle &\scriptstyle i&\scriptstyle &\scriptstyle 1&\scriptstyle -i&\scriptstyle
\\\scriptstyle1&\scriptstyle &\scriptstyle &\scriptstyle 1&\scriptstyle &\scriptstyle 1&\scriptstyle -i&\scriptstyle &\scriptstyle &\scriptstyle i&\scriptstyle 1&\scriptstyle &\scriptstyle 1&\scriptstyle &\scriptstyle &\scriptstyle 1
\\\scriptstyle i&\scriptstyle &\scriptstyle &\scriptstyle -i&\scriptstyle &\scriptstyle -i&\scriptstyle 1&\scriptstyle &\scriptstyle &\scriptstyle -1&\scriptstyle -i&\scriptstyle &\scriptstyle -i&\scriptstyle &\scriptstyle &\scriptstyle -i
\\&\scriptstyle -1&\scriptstyle i&\scriptstyle &\scriptstyle -1&\scriptstyle &\scriptstyle &\scriptstyle -1&\scriptstyle i&\scriptstyle &\scriptstyle &\scriptstyle i&\scriptstyle &\scriptstyle 1&\scriptstyle -i&\scriptstyle
\\&\scriptstyle -i&\scriptstyle -1&\scriptstyle &\scriptstyle -i&\scriptstyle &\scriptstyle &\scriptstyle -i&\scriptstyle -1&\scriptstyle &\scriptstyle &\scriptstyle -1&\scriptstyle &\scriptstyle i&\scriptstyle 1&\scriptstyle
\\\scriptstyle-i&\scriptstyle &\scriptstyle &\scriptstyle -i&\scriptstyle &\scriptstyle -i&\scriptstyle -1&\scriptstyle &\scriptstyle &\scriptstyle 1&\scriptstyle -i&\scriptstyle &\scriptstyle -i&\scriptstyle &\scriptstyle &\scriptstyle -i
\\\scriptstyle1&\scriptstyle &\scriptstyle &\scriptstyle 1&\scriptstyle &\scriptstyle 1&\scriptstyle -i&\scriptstyle &\scriptstyle &\scriptstyle i&\scriptstyle 1&\scriptstyle &\scriptstyle 1&\scriptstyle &\scriptstyle &\scriptstyle 1
\\&\scriptstyle -i&\scriptstyle -1&\scriptstyle &\scriptstyle -i&\scriptstyle &\scriptstyle &\scriptstyle -i&\scriptstyle -1&\scriptstyle &\scriptstyle &\scriptstyle -1&\scriptstyle &\scriptstyle i&\scriptstyle 1&\scriptstyle
\\\scriptstyle1&\scriptstyle &\scriptstyle &\scriptstyle 1&\scriptstyle &\scriptstyle 1&\scriptstyle -i&\scriptstyle &\scriptstyle &\scriptstyle i&\scriptstyle 1&\scriptstyle &\scriptstyle 1&\scriptstyle &\scriptstyle &\scriptstyle 1
\\&\scriptstyle 1&\scriptstyle -i&\scriptstyle &\scriptstyle 1&\scriptstyle &\scriptstyle &\scriptstyle 1&\scriptstyle -i&\scriptstyle &\scriptstyle &\scriptstyle -i&\scriptstyle &\scriptstyle -1&\scriptstyle i&\scriptstyle
\\&\scriptstyle i&\scriptstyle 1&\scriptstyle &\scriptstyle i&\scriptstyle &\scriptstyle &\scriptstyle i&\scriptstyle 1&\scriptstyle &\scriptstyle &\scriptstyle 1&\scriptstyle &\scriptstyle -i&\scriptstyle -1&\scriptstyle
\\\scriptstyle1&\scriptstyle &\scriptstyle &\scriptstyle 1&\scriptstyle &\scriptstyle 1&\scriptstyle -i&\scriptstyle &\scriptstyle &\scriptstyle i&\scriptstyle 1&\scriptstyle &\scriptstyle 1&\scriptstyle &\scriptstyle &\scriptstyle 1
\end{array}\right],
\label{sfgpayoff}
\end{equation}
while the general strategy state when there is no cooperation
between players is
\begin{equation}
\rho^{q,S} = \left[\begin{array}{llll}\scriptstyle\rho^{q,1}_{ii}
& \scriptstyle\rho^{q,1}_{ix} & \scriptstyle\rho^{q,1}_{iy} &
\scriptstyle\rho^{q,1}_{iz}
\\
\scriptstyle\rho^{q,1}_{xi} & \scriptstyle\rho^{q,1}_{xx} &
\scriptstyle\rho^{q,1}_{xy} & \scriptstyle\rho^{q,1}_{xz}
\\
\scriptstyle\rho^{q,1}_{yi} & \scriptstyle\rho^{q,1}_{yx} &
\scriptstyle\rho^{q,1}_{yy} & \scriptstyle\rho^{q,1}_{yz}
\\\scriptstyle\rho^{q,1}_{zi} & \scriptstyle\rho^{q,1}_{zx} &
\scriptstyle\rho^{q,1}_{zy} & \scriptstyle\rho^{q,1}_{zz}
\end{array}\right]\otimes\left[\begin{array}{llll}\scriptstyle\rho^{q,2}_{ii} & \scriptstyle\rho^{q,2}_{ix} &
\scriptstyle\rho^{q,2}_{iy} & \scriptstyle\rho^{q,2}_{iz}
\\
\scriptstyle\rho^{q,2}_{xi} & \scriptstyle\rho^{q,2}_{xx} &
\scriptstyle\rho^{q,2}_{xy} & \scriptstyle\rho^{q,2}_{xz}
\\
\scriptstyle\rho^{q,2}_{yi} & \scriptstyle\rho^{q,2}_{yx} &
\scriptstyle\rho^{q,2}_{yy} & \scriptstyle\rho^{q,2}_{yz}
\\\scriptstyle\rho^{q,2}_{zi} & \scriptstyle\rho^{q,2}_{zx} &
\scriptstyle\rho^{q,2}_{zy} & \scriptstyle\rho^{q,2}_{zz}
\end{array}\right].
\label{sfgstate}
\end{equation}
Compare equ(\ref{sfgpayoff}) and equ(\ref{sfgstate}) with
equ(\ref{pfgpayoff}) and equ(\ref{pfgstate}), $16\times16$
full-structure matrices are used to replace $4\times4$ diagonal
matrices, first, because we have four base vectors of strategy
space in SFG other than two in PFG; second, because of the
non-zero off-diagonal elements. Such elements have no
corresponding meaning in classical game: what's the meaning of
$\left<XX\right|H^{1}\left|II\right> = 1$? From left side, it
looks like both players choose $X$, while from right side, both
players choose $I$!

In fact, this remind us the meaning of the off-diagonal terms in
density matrix and Hamiltonian of a quantum object. There such
terms also have no classical correspondence, and they are a
distinguishable character of quantum system compared with
classical system. Therefore, the relation between TGT and QGT
looks exactly like the relation between Classical Mechanics and
Quantum Mechanics.

\section{Discussion}
\label{conclusion}

In fact, the idea of Quantum Game Theory has long been proposed in
\cite{meyer} and developed in \cite{jens, basevec}, and currently
in an active development stage\cite{abbott}. However, in all the
former general prescription of Quantum Game Theory, although they
do consider games on quantum objects, the strategy state is always
treated as a probability distribution function over
$\mathcal{H}^{*}$, the whole operator space, not a density matrix
expanded on $\mathbb{B}\left(\mathcal{H}^{*}\right)$, a basis of
$\mathcal{H}^{*}$. Obviously, those two descriptions of state are
different. In \cite{wugame}, we did a detailed comparison between
them and gave an argument that why our density matrix
representation should be used instead of the probability
distribution function.

In this paper, we first related abstract strategies in Game Theory
with operators acting on physical objects, named game objects.
Then, operators are treated as vectors in Hilbert space. Because
Traditional Game Theory use probability distribution functions
over the classical operators Hilbert space to act as general
mixture strategy, the density matrix form of a probability
distribution function is diagonal. However, in Quantum Game
Theory, when we replace the classical game object with a quantum
object, a strategy state must be a full-structured density matrix
over Hilbert space of quantum operators, not the diagonal density
matrix coming from probability distribution function. This is just
like the relation between Classical Mechanics and Quantum
Mechanics.

Besides the non-cooperative game, this new framework of Game
Theory can also be used to discuss Coalitional Game Theory (CGT).
When the system-level density matrix,
$\rho^{S}\neq\prod_{i}\rho^{i}$, not a direct-product state, it
naturally leads to correlation between players. This implies
cooperations between players. The possibility to link this new
framework with CGT will be an interesting topic\cite{wugame}.

\section{Acknowledgement}
Thanks Dr. Shouyong Pei for the stimulating discussion during
every step of progress of this work.

\end{document}